# Supramolecular self-assembly as a tool to preserve electronic purity of perylene diimide chromophores

Ina Heckelmann[1,2], Zifei Lu[3], Joseph C. A. Prentice[4], Florian Auras[1], Tanya K. Ronson[3], Richard H. Friend[1], Jonathan R. Nitschke[3], Sascha Feldmann[1,5*]

[1]*Cavendish Laboratory, University of Cambridge, Cambridge, CB30HE, UK*
[2]*Institute for Quantum Electronics, ETH Zürich, Zurich, CH-8093, Switzerland*
[3]*Yusuf Hamied Department of Chemistry, University of Cambridge, Cambridge, CB21EW, UK*
[4]*Department of Materials, University of Oxford, Oxford, OX13PH, UK*
[5]*Rowland Institute, Harvard University, Cambridge, MA-02142, USA*
[*]*Email: sfeldmann@fas.harvard.edu*

**Abstract**

Small molecule organic semiconductors hold great promise for efficient, printable, and flexible optoelectronic applications like solar cells and displays. However, strong excited-state quenching due to uncontrolled aggregation currently limits their performance and employability in devices. Here, we report on the self-assembly of a supramolecular pseudo-cube formed from six modified tetradentate perylene diimides (PDIs). The rigid, shape-persistent cage sets the distance and orientation of the PDI chromophores and suppresses intramolecular rotations and vibrations, leading to non-aggregated, monomer-like electronic properties in solution as well as in the solid state, in contrast to the fast fluorescence quenching in the free ligand. The stabilized excited state and electronic purity of the cage enable the observation of delayed fluorescence due to a bright excited multimer state, which acts as an excited state reservoir, due to a rare case of benign inter-chromophore interactions in the cage. Our results suggest that not only the photophysical properties of the subcomponents but the geometric structure is crucial for the overall optoelectronic properties of supramolecular systems. We show that self-assembly provides a powerful tool for retaining and controlling the electronic properties of well-studied chromophores, providing a route to bring molecular electronics applications in reach.

**Introduction**

π-conjugated chromophores have been known for centuries and are omnipresent, for example as photo-active molecules in photosynthesis or simply when used to dye fabrics for clothing[1,2]. More recently, such dye molecules have been intensively studied for their optoelectronic properties in organic and molecular electronics applications, for example for organic photovoltaics (OPV) or organic light-emitting diodes (OLEDs), now commonly found in many displays when seeking high brightness and color-purity[3,4]. Perylene diimide (PDI) represents such a prototypical chromophore molecule often used as structural motif for pigments or molecular electronics[5,6].

However, PDI, like the vast majority of highly conjugated planar aromatic compounds suffers from strong excited-state quenching and loss of its favorable electronic properties upon aggregation through π-π stacking, for example in concentrated solutions or thin films, or due to rovibrational losses[7–9]. Such undesired inter-chromophore interactions because of uncontrolled supramolecular assembly present one of the largest challenges yet to be overcome before organic molecules can replace the currently dominating crystalline inorganic semiconductors like silicon or gallium arsenide as future technologies based on plastic electronics.

Here we report a multi-chromophore system assembled in a fully controlled fashion using synthetic supramolecular chemistry based on coordination complexes. We design a pseudo-cube consisting of six PDI-derivative molecules which self-assemble into a cage-like structure. The rigid framework of the cage retains the electronic purity of the PDI chromophore, which is otherwise lost in the uncomplexed PDI ligands due to vibrational losses or π-stacking. The rich photophysics of the cage's excited state, now stabilized through restriction of intramolecular rotation, reveals rarely observed benign inter-chromophore interactions, resulting in a hitherto unobserved emissive excited multimer state that is in equilibrium with the excited singlet to yield delayed fluorescence of the cage. Our study shows how supramolecular self-assembly provides a new platform to retain and control the electronic properties of well-studied chromophores, providing a powerful route to bring molecular electronics applications in reach.

**Results**

We first synthesize a suitably modified building block for subsequent supramolecular self-assembly based on the prototypical chromophore perylenediimide (PDI) by attaching phenylamine groups to the bay area and neopentyl groups to the head area for enhanced solubility (see SI for details on precursor synthesis). The thus modified PDI chromophore ligand **L1** reacted with zinc(II) ions and picolinal to yield the supramolecular pseudo-cube **C1**, formed from 6 such ligands (see Fig. 1).

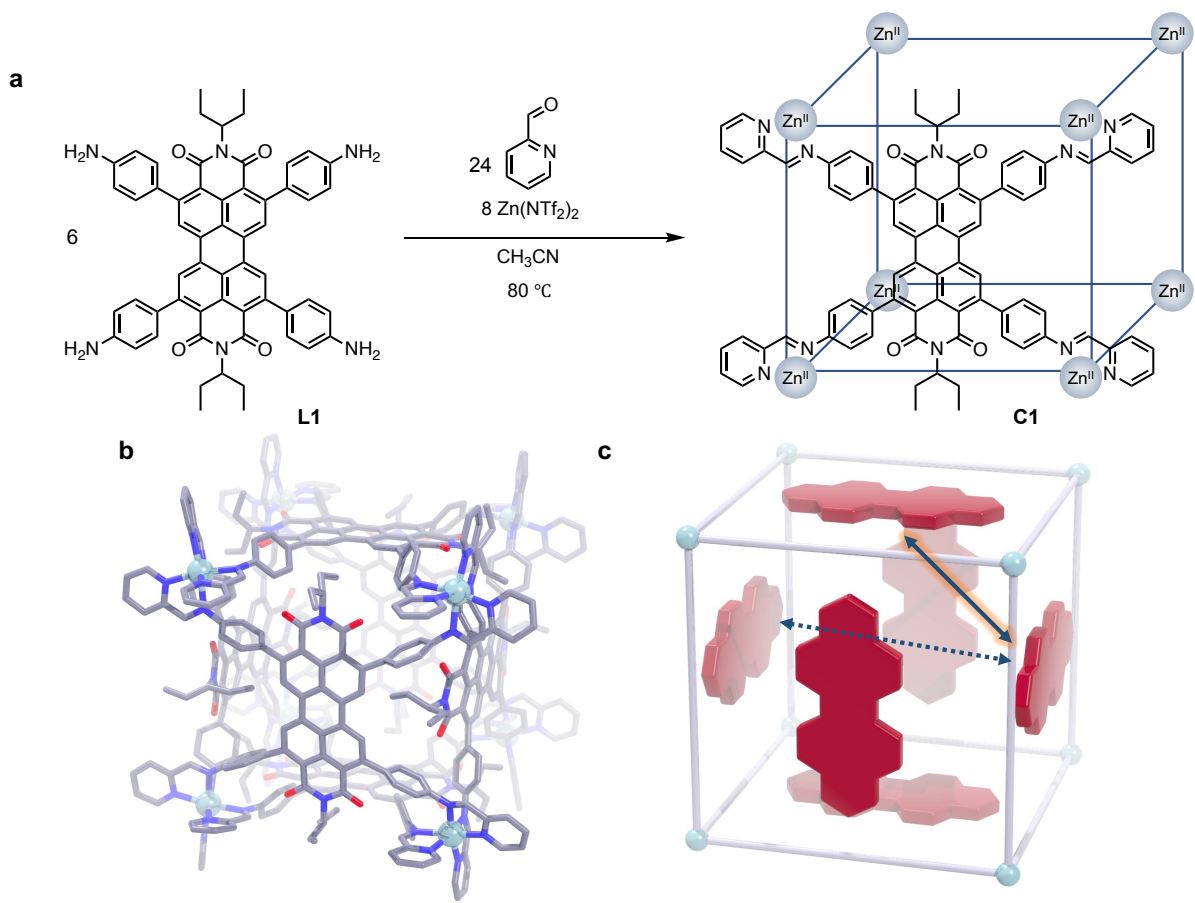

**Fig. 1 | Synthesis and structure of the supramolecular pseudo-cube. a**, Self-assembly of PDI ligand **L1** into pseudo-cube **C1** with only one of six ligands depicted for clarity. **b**, Structure of **C1** as derived from synchrotron-based single-crystal X-ray diffraction. Counter-ions and co-crystallized solvent molecules are omitted for clarity. **c**, Relative orientation of possible PDI transition dipole moment couplings in the cubic cage.

We confirmed the cube-like cage structure by synchrotron-based single-crystal X-ray diffraction (Fig. 1b). Importantly, the question arises to what extent this controlled supramolecular self-assembly impacts the optoelectronic properties of the PDI chromophore compared to the free ligand: Both H- and J-type exciton interactions could be hypothesized to occur, which could be favorable or unfavorable with respect to the photoluminescence (PL) of the resulting complex (Fig. 1c).

We next investigate the steady-state optical properties of the cube-like PDI cage **C1** and compare them with those of the free ligand **L1** (Fig. 2).

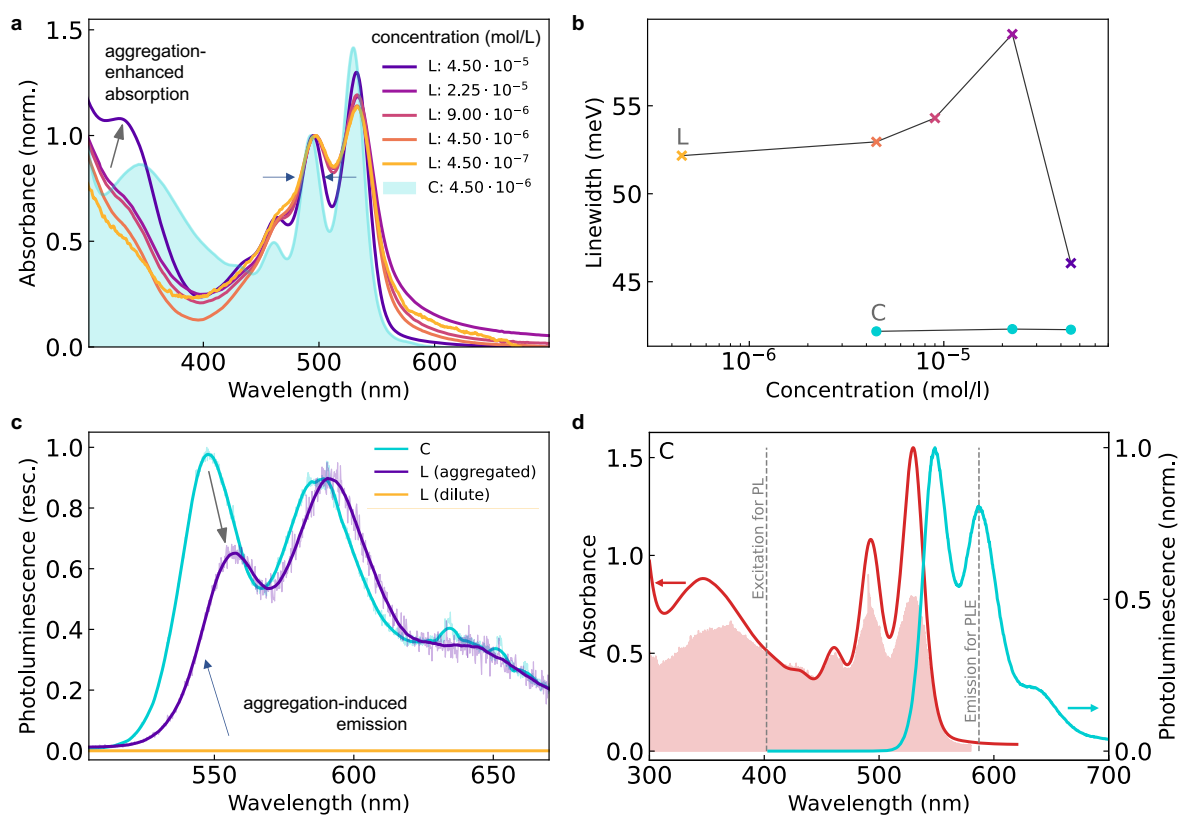

**Fig. 2 | Supramolecular self-assembly as a tool for the controlled restoration of the electronic purity of PDI. a**, Concentration-dependent absorbance of **L1** with retrieval of sharp absorption peaks above a threshold concentration, when also an additional feature at 330 nm is observed, indicating aggregation. In contrast, the cage assembly **C1** (filled cyan area) shows significantly lower energetic disorder. **b**, Concentration-dependent 0-0 absorption peak linewidth of free ligand **L1** and its cubic assembly **C1**, quantifying the onset of uncontrolled aggregation in the ligand in contrast to the unchanging high electronic order retained in **C1**. **c**, Photoluminescence (PL) of dilute free ligand (yellow, fully quenched), highly concentrated aggregated ligand (purple, weak emission), and cage assembly (cyan, stronger emission like from pure PDI). Excited at 405 nm. **d**, Direct comparison between absorbance (red line), PL (cyan line) and PL excitation scan (PLE, red filled area) shows the narrow bands, clear mirror-symmetry, and high electronic purity of the PDI achieved in the cage assembly.

For the free PDI ligand **L1**, we find significant electronic disorder leading to broadened absorbance features in its vibronic progression (Fig. 2a). This energetic disorder initially increases upon increasing its concentration until at a certain threshold concentration the linewidth suddenly drops (see also Fig. 2b), in conjunction with the occurrence of an additional absorbance feature around 330 nm. This is indicative of the onset of spontaneous aggregation of the ligand in solution. In contrast, we observe for the cage assembly **C1** a very narrow linewidth throughout all tested concentrations, implying a high degree of electronic purity is achieved through the restriction of the PDI ligands' motion once confined as part of the cage. We see the far-reaching consequences of this energetic disorder in the free PDI ligand with respect to its emission properties (Fig. 2c): Its photoluminescence (PL) is fully quenched (yellow line) with a PL quantum yield (PLQY) below our detection limit of $10^{-4}$ (see SI for details). We discuss the origin of this quenching in the context of rovibrational losses further below. Only upon aggregation at high concentration is the PL partially restored (purple line, PLQY ca. 0.1%).

In stark contrast to these observations, for the cage self-assembly (cyan line) a much stronger PL observed, showing an (about 100-fold) increased PL efficiency, even at low concentrations where electronic disorder and quenching prevails for the ligand. This preserved electronic purity of the PDI emission in the cage as opposed to that in the free PDI ligand becomes even more obvious when directly comparing the cage absorbance and PL spectra (Fig. 2d), showing near-perfect mirror symmetry between the spectra. For the uncontrolled PDI ligand aggregation (Fig. 2c), the PL spectrum is instead truncated in its 0-0 peak and is indicative of H-type aggregation[10]. The PL excitation scan (PLE, monitored at the emission center at 585 nm) of the cage also confirms the absence of any non-PDI core states that one could argue would influence the optoelectronic properties of the PDI upon self-assembly into the cage. Similarly, no additional low-energy absorbance tail is observed for the cage compared to a pure PDI molecule, confirming the absence of any unwanted extended conjugation between the PDI chromophores in the cage assembly that would impact its electronic purity. We also confirm that at higher concentrations the cage retains these electronically pure properties, and we show that changes to the PL spectrum at high concentrations can be described purely by accounting for self-absorption (see SI Fig. S2), in contrast to the PL changes observed for the pure ligand upon uncontrolled aggregation, mentioned above.

Supramolecular self-assembly therefore serves here as a powerful tool to retain the electronic purity of its subcomponents which would otherwise be lost when trying to functionalize a chromophore to make it useful for optoelectronic applications, for example shown in the restoration of the PL otherwise lost in the ligand or only partially restored *via* its uncontrolled aggregation.

Having confirmed this, we now explore the excitonic properties of the pseudo-cubic chromophore cage compared to the free PDI ligand through time-resolved PL spectroscopy (Fig. 3).

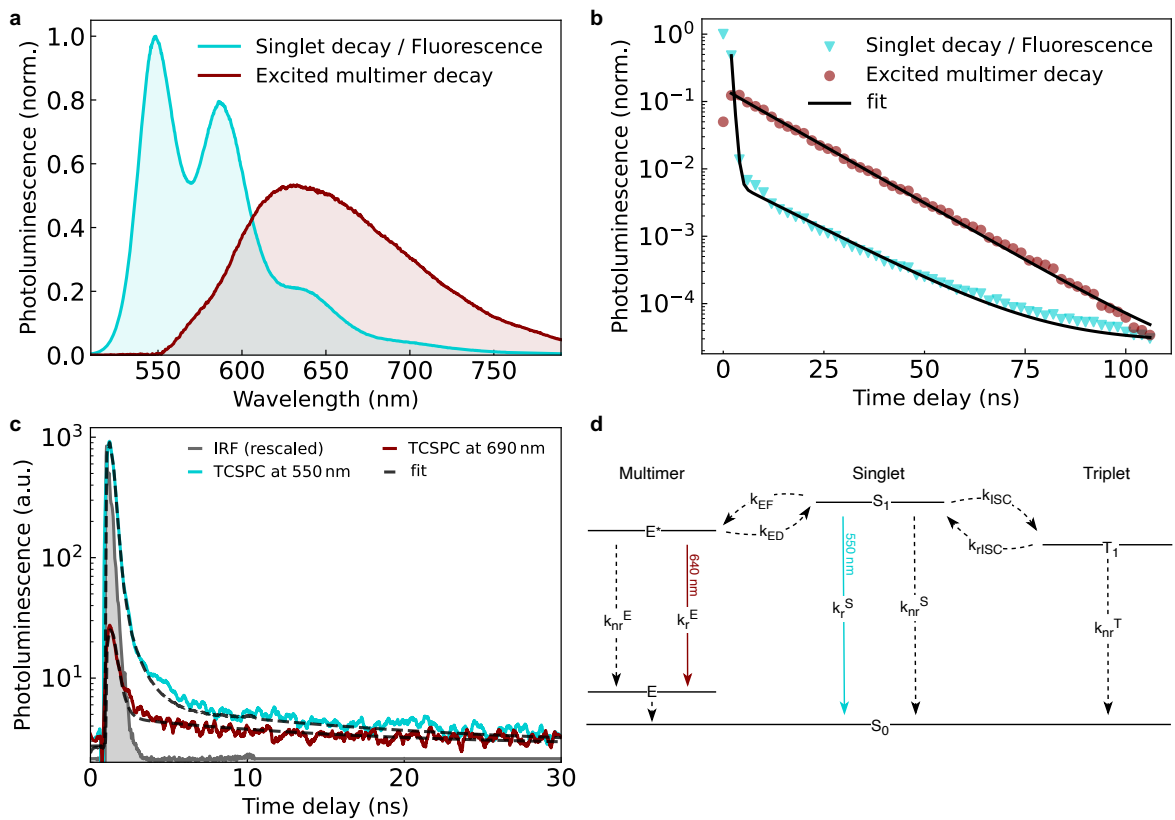

**Fig. 3 |** Observation of a bright excited multimer and resulting delayed fluorescence in a supramolecular self-assembly of chromophores. **a**, Prompt singlet (cyan) and delayed excited multimer (red) emission of the PDI pseudo-cube **C1** measured upon excitation at 400 nm (~100 fs pulsed, repetition rate 1 kHz) using nanosecond electrical gating (see SI for setup details). **b**, Kinetics of the spectral components shown in **a** with lifetime fits (see SI for fitting details and rate constants). The singlet decay shows typical delayed fluorescence characteristics through re-population by a dark triplet and bright excited multimer state acting as excitation reservoirs. The excited multimer emission shows a monoexponential lifetime of 12.8±0.2 ns. **c**, Time-correlated single-photon counting (TCSPC)-based kinetics of the singlet and excited multimer state upon excitation with 405 nm laser pulses. The higher time-resolution and convolution of the fits with the instrument response function, measured by excitation-laser scattering off a glass colloid (grey area), reveals the faster lifetimes involved in the multi-state system (see SI for details and underlying fitting model). **d**, Jablonski diagram summarising all energy levels and interconversion processes observed in the supramolecular PDI cube as determined from transient PL and transient absorption (TA) measurements; E*, unrelaxed and E, relaxed excited multimer, S, singlet, T, triplet, $k_r$ and $k_{nr}$ radiative and non-radiative recombination rate of each species, respectively.

In contrast to the fully quenched PL of the free ligand, we can observe rich photophysics in the excited state of the supramolecular PDI cube. This is a direct consequence of the strong restriction of intramolecular rotation in the cage-like self-assembly, whereas the free rotation of the phenylamine groups in the case of the free ligands leads to fast non-radiative decay to the ground state upon photoexcitation. Surprisingly, upon time-gating we not only observe an expected bright singlet emission of the cage, but also a spectrally well separated red-shifted and broad emission band centered around 640 nm (Fig. 3a). This is strongly reminiscent of the excimer emission observed before for covalently bound PDI-dimers and oligomers, for example by the groups of Wasielewski (*e.g.* refs.[11,12]) or Würthner (*e.g.* ref.[13] and citations therein). Importantly, this species forms only upon photo-excitation and subsequent electronic coupling of the chromophores with each other, followed by structural reorientation, explaining why this state is not observed in the steady-state absorbance spectrum shown earlier.

This observation is impressive in two ways: i) It is one of the very few examples of benign (favorable) rather than malign (unfavorable) exciton-exciton interactions observed in a multi-chromophore system, where emission usually is fully quenched[14–16] – we, for example, reported on full PL quenching in a chromophore-based tetrahedral cage[17] – rather than the 100-fold increase of singlet emission we observe in the present case. ii) We observe, to the best of our knowledge, for the first time an emissive excited multimer state in a supramolecular self-assembly of dye molecules, as a consequence of the preservation of the excited state due to the restriction of intramolecular rotational losses and the correct chromophore orientation in space, allowing for this favorable inter-chromophore coupling to take place. As we cannot determine directly the exact number of coupled chromophores in our system, we refer to this new state as *excited multimer* (E* upon photo-induced formation, and structurally relaxing into state E), rather than as *excimer* (implying typically excited-state dimer, *i.e.* limited to only two chromophores).

Moreover, the kinetics of these excited states reveal a distinct delayed emission component for the singlet (Fig. 3b-c), resembling those observed in thermally-activated delayed fluorescence emitters[18–20] (see SI for modelling details). This is the result of the crossing and reverse-crossing between the bright singlet, the bright excited multimer, and – as we confirm below using transient absorption (TA) spectroscopy – a dark triplet state (Fig. 3d). Both the excited multimer and the triplet are needed to explain all observed kinetics, and both can act as excitation reservoirs re-populating the singlet in a dynamic equilibrium, although we note that the majority of excitons in the self-assembly still ultimately decay non-radiatively, which we rationalize below using insights from TA and density functional theory (DFT). We also highlight the absence of any low-energy absorption onset for the cage UV-vis spectrum (Fig. 1d), further excluding the formation of any aggregates and related state hybridization in the cage's electronic ground state.

We now confirm these photophysical processes in the cage and their absence in the free ligand by employing TA spectroscopy, which allows us to also resolve non-emissive species (Fig. 4).

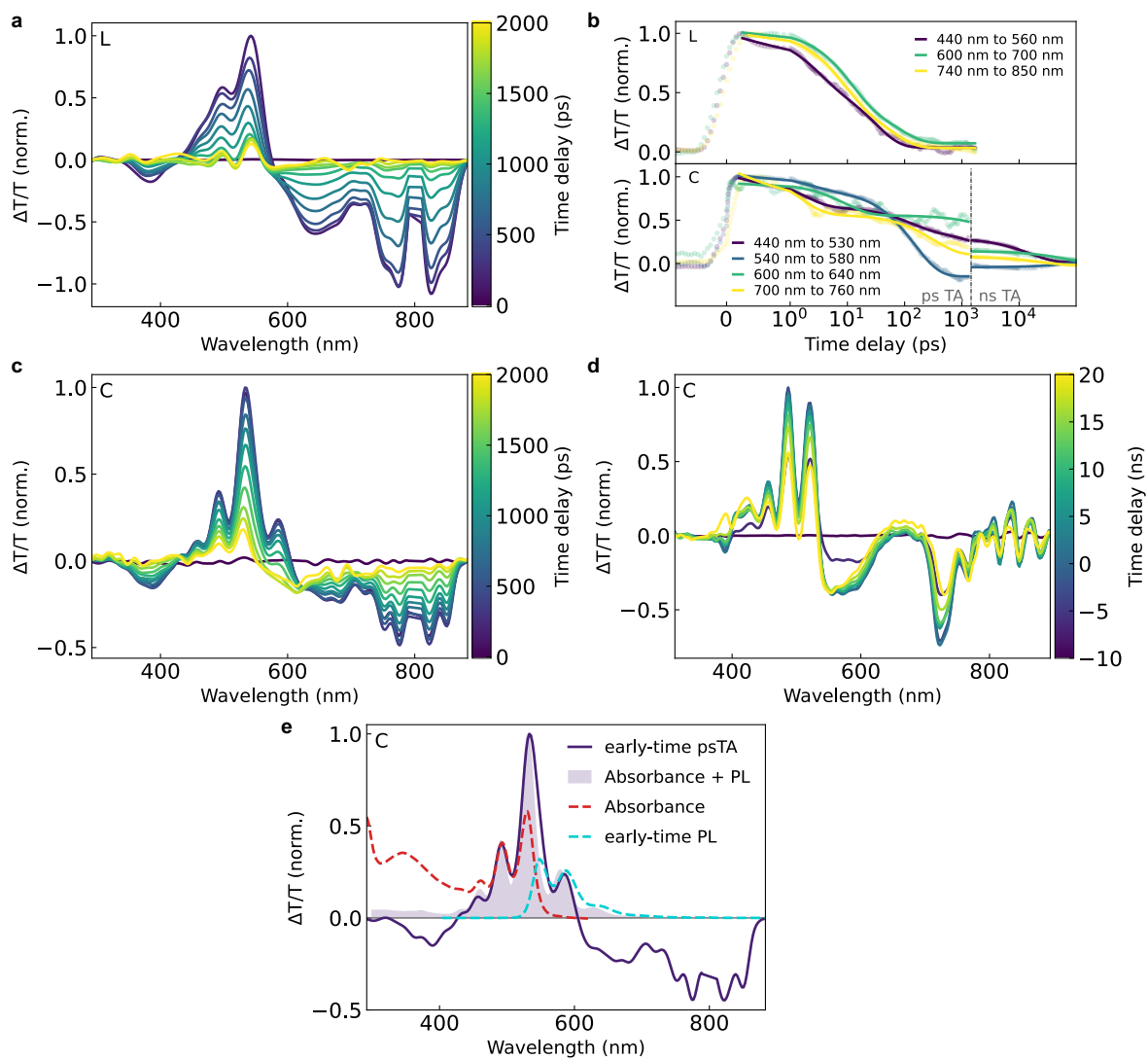

**Fig. 4 | Confirmation of excited multimer and triplet dynamics in supramolecular PDI cage with transient absorption (TA) spectroscopy**, unravelling the rare case of benign inter-chromophore interaction in a supramolecular assembly, where the triplet and excited multimer states act as reservoirs for delayed luminescence. **a**, ps-TA spectra of PDI ligand **L1**, and **b**, kinetics showing excited-state quenching with a lifetime of 18 ps for the free ligand, but largely extended lifetimes for the cage. We note that the spectral indent at 800 nm is due to residual 800 nm laser scatter which is used to generate pump and probe beams in the lab. **c**, ps- and **d**, ns-TA spectra for the PDI cage **C1**, showing an extended excited-state lifetime into the ns-range and the emergence of the positive excited-multimer stimulated-emission band around 650 nm at later times for the cage. **e**, the early-time TA spectrum of the PDI cage can be well reconstructed from the contributions from its absorbance (ground-state bleach) and early-time PL (stimulated emission), the latter of which is absent in the ligand TA spectrum when comparing **a** and **c**, as all emission is quenched for the ligand. See SI for experimental details and fitting procedure.
8

The transient absorption data shows the changes in absorbance of the material after initial excitation with a pulsed pump beam (400 nm for the ultrafast measurements) and mapped across different time delays between the pump and the broadband probe beam, measured in transmission (*T*). For the pure PDI ligand in solution (Fig. 4a), we find a positive change in transmission signal from around 450-570 nm which can be assigned to the ground-state bleach (GSB) of the $S_0$ to $S_1$ transition, and a broad negative signal from about 600 nm onwards, which corresponds to overlapping photo-induced absorption (PIA) peaks from the $S_1$ state into higher lying states and, upon ultrafast intersystem crossing, from $T_1$ into higher lying states[21,22]. Figure 4b shows spectrally integrated kinetics of these transitions, revealing that the entirety of the excited state population of the free ligand decays back to the ground state with an average monoexponential lifetime of 18 ps, much faster than the nanosecond radiative lifetime PDI, explaining why the PL of the ligand is fully quenched.

In contrast, for the supramolecular cage assembly the excited-state lifetime extends far into the ns-regime, in line with the rich emission properties discussed above. For the PDI cage, the TA spectrum (Fig. 4c,d) also shows modified spectral features in the region of the vibronic progression of the GSB, as in this bright emitter stimulated emission (SE) can be observed as a positive change in transmission signal at early times. Figure 4e demonstrates how the early-time TA spectrum of the cage can be well understood in terms of a superposition of its absorbance and emission properties. At ns-time delays (not measurable for the PDI ligand which is fully relaxed back to the ground state by then) for the cage we moreover see a broad positive signal around 650 nm (Fig. 4d), which matches well the excited multimer emission observed in the PL experiments. We anticipate this to be a SE feature of the excited multimer convoluted with the PIA (now decreased in intensity) in this region – as this excited multimer state only forms delayed after photo-excitation it explains why this signal is absent in the early-time TA data. Aside from the fact that the triplet state in PDI molecules is well reported in the literature[21], we also note that in order to fit all kinetics of the PDI cage (*i.e.* ns-iCCD & ps-TCSPSC PL data and ns- & fs/ps-TA data) we do need to include both the dark triplet state and the bright excited multimer state in the model (see SI for details).

To further understand our experimental results, we performed first principles calculations, examining the excitations of the PDI ligand with its phenylamine linkers using time-dependent density functional theory (TDDFT). In particular, we investigated the excitations for several dihedral angles between the phenyl-ring moieties and the PDI core (Fig. 5).

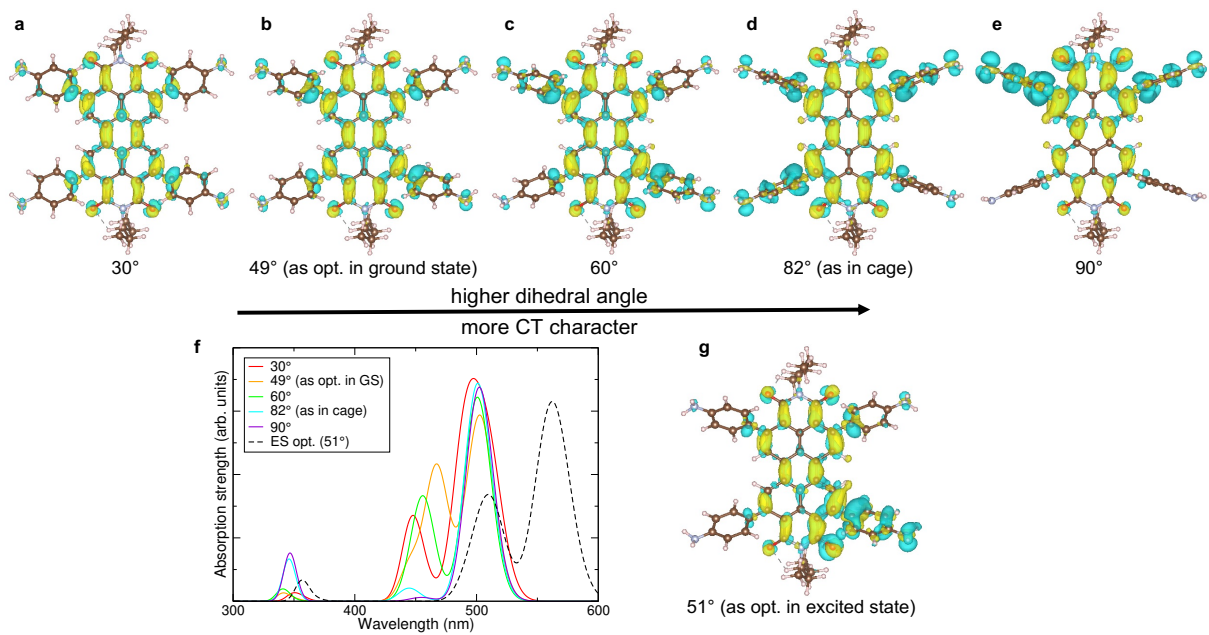

**Fig. 5 | Restriction of intermolecular rotation in cage supported by first principles. a-e,g**, Charge density distributions of the PDI ligand **L1** calculated for various degrees of phenyl-ring out-of-plane rotations, including ground-state structure as optimized in DFT (**b**), the phenyl-ring angle seen in the cage (**d**), and the excited-state structure as optimized in TDDFT (**g**). **f**, Absorbance spectra for the various configurations. Excitations strongly couple to the dihedral angle between the peripheral phenylamine rings and the PDI core, with higher angles leading to stronger charge transfer (CT) character (**a-e**). In the excited state, the peripheral phenylamines adopt a more twisted configuration, leading to increased CT character (**f,g**). Along with the strong coupling to ligand rotation, this leads to quenching of emission in the free ligand molecule. This is then prevented through the rigid framework in the cage assembly – thus the pure electronic PDI properties are retained.

We calculated the absorption spectra of the PDI ligand **L1**, using the highly accurate OT-LRC-$\omega$PBE functional[23,24], as a function of the rotation angle of the phenylamine rings, (Fig. 5f). The positions of the two strongest absorption peaks at around 450 and 500 nm compare very well with the experimental absorption data shown earlier in Fig. 2. The peak at around 500 nm is largely unaffected by the dihedral angle, whereas the peak at around 450 nm dramatically decreases in strength as the angle increases, implying a strong coupling to the rotation of the phenylamines. This coupling is further explored by examining the difference density for the 450 nm excitation at the various dihedral angles, as shown in Fig. 5a-e. This shows that the character of the excitation is also strongly dependent on the dihedral angle. At low angles, the excitation is largely localised on the PDI core, but at higher angles, the phenylamine rests become more involved, acting as electron donors, with the excitation taking on a stronger charge transfer character. This charge transfer character makes radiative recombination less likely, and correlates with the decrease in the oscillator strength with increasing dihedral angle, as seen in Fig. 5f.

We further quantify the coupling of this excitation with the rotation of the ligands by calculating the non-adiabatic coupling vector (NACR), which provides a measure of the coupling between electronic excitation and atomic motion. We calculated the coupling between the excitation at around 450 nm and two sets of atomic motion: i) the rotation of a ligand, and ii) a breathing mode of the PDI core, which is also relevant to the relaxation from the equilibrium geometry in the ground state to that in the excited state (see SI for computational details). We find that the coupling of the electronic excitation to the rotation of the ligands is 6.7 times stronger than to the PDI breathing mode, which further demonstrates that there is a strong coupling between the excitation and the rotation of the ligands, responsible for the non-radiative losses in the free molecule. Restriction of this rotation in the supramolecular cage thus explains the observed enhancement in luminescence compared to the free PDI ligand.

Finally, we also optimize the geometry of the system in the excited state corresponding to the 450 nm excitation, and recompute the absorption spectrum and difference density for this configuration (Fig. 5f-g). Importantly, the excited state equilibrium geometry has a larger dihedral angle between the phenylamine ligands and the PDI core than the ground state equilibrium geometry (49° *vs*. 51°). We expect the change in angle to be somewhat underestimated, as the PBE functional, which has known issues with describing charge transfer, was used for the excited state geometry optimisation for reasons of computational cost. Apart from a global red-shift of around 50 nm, the position, strength, and character of the '450 nm' excitation in the excited state equilibrium geometry match well with what would be expected from the results of pure rotations of the ligands.

Our calculated energetic positions of the $S_1$ state at 2.46 eV and the $T_1$ state at 1.18 eV for the free PDI ligand, respectively (2.47 eV and 1.13 eV in the cage geometry for each case) also match very well with the positions determined experimentally from the PL (singlet feature) and TA (singlet and triplet features).

Taken all together, the computational results imply that electronic excitations couple strongly to the rotation of the phenylamine ligands relative to the PDI core, providing a route for non-radiative energy loss and recombination. Simultaneously, exciting the system leads to the dihedral angle relaxing towards a larger dihedral angle, which is associated with a decrease in oscillator strength, making radiative recombination less likely. This provides an explanation for the quenching of photoluminescence in the free molecule. In the cage, however, the ligands are not free to rotate. The ground state dihedral angle between the PDI core and phenylamine ligands is much larger in the cage than in the free molecule, meaning that the oscillator strength is weak (though non-zero), but non-radiative recombination via coupling to ligand rotation is no longer possible. This allows radiative recombination to become competitive, producing the observed increase in photoluminescence and extending the excited state lifetime to allow for the inter-chromophore coupling induced excited multimer emission to be observed in the supramolecular cage.

We also note that the out-of-plane twisted configuration of the phenyl rings with respect to the PDI core result in a more pronounced CT character of the excited state for the PDI ligand, which would in principle facilitate inter-system crossing to form triplets, reportedly on ultrafast timescales. This would explain the loss of PL we observe for the PDI ligand, but would not explain the fast excited state decay within 18 ps that we observe in TA (the triplet state would be long-lived). Instead, due to the strong non-radiative losses resulting from the coupling to the ring rotations, all excitations are ultimately quenched non-radiatively in the case of the free ligand, as also supported by our calculations. In contrast, the restriction of intramolecular rotation in the cage assembly leads to the successful recovery of the intrinsic optical properties expected from the pure PDI ligand, *i.e.* it shows both intersystem-crossing to form triplets, emission from its excited singlet state (which has still significant CT character and reduced oscillator strength as our calculations also confirm, but an at least 100-fold increase in PL compared to the freely rotating ligand), and, importantly an excited multimer emissive state which acts together with the triplet state as an excitation reservoir for observing delayed emission.

**Conclusions**

In summary, we report on the synthesis of a supramolecular pseudo-cube formed from modified PDI dye molecules. We find this cage assembly to show 100-fold enhanced emission compared to its free PDI ligand subcomponent and that the rigid cage framework acts as a powerful tool to retain the electronic purity of the chromophores which is typically lost in uncontrolled aggregation. The suppression of non-radiative vibrational losses allows for the observation of an emissive excited multimer state through benign inter-chromophore coupling enabled by the high level of assembly control, paving the way for tailored optoelectronic applications following chemical bottom-up design.

**Corresponding Author**

*Sascha Feldmann** – *Cavendish Laboratory, University of Cambridge, Cambridge, CB3 0HE, United Kingdom &* Rowland Institute, Harvard University, Cambridge, MA 02142, USA

Email: sfeldmann@fas.harvard.edu

**Acknowledgements**

I.H. acknowledges funding from the Studienstiftung des deutschen Volkes. Z.L. acknowledges PhD funding from the Cambridge Trust and the China Scholarship Council. J.C.A.P. acknowledges the support of St Edmund Hall, University of Oxford, through the Cooksey Early Career Teaching and Research Fellowship. We are grateful for computational support from the UK national high performance computing service, ARCHER2, for which access was obtained via the UKCP consortium and funded by the Engineering and Physical Sciences Research Council (EPSRC UK) *via* grant ref EP/P022561/1. F.A. acknowledges funding from the European Research Council (ERC) under the European Union's Horizon 2020 research and innovation programme (grant agreement No 670405).

This study was supported by the European Research Council (grant 695009) and the EPSRC UK grant EP/P027067/1. The authors also thank Diamond Light Source (UK) for synchrotron beamtime on I19 (CY21497), and the NMR service in the Yusuf Hamied Department of Chemistry at the University of Cambridge for NMR experiments. S.F. acknowledges funding from the EPSRC UK *via* an EPSRC Doctoral Prize Fellowship, and from the Rowland Institute at Harvard University. S.F is grateful for support from the Winton Programme for the Physics of Sustainability.

**NOTES**

The authors declare no competing interest.